\documentclass[useAMS,usenatbib]{mn2e}
\usepackage{graphicx}
\title[Radial-velocity study of the CV SDSS\,J1433+1011]{Radial-velocity study of the post-period minimum cataclysmic
variable SDSS\,J143317.78+101123.3 with an electron-multiplying CCD}
\author[S. M. Tulloch, P. Rodr\'{i}guez-Gil and V. S. Dhillon]{S. M. Tulloch$^{1,2}$\thanks{E-mail: smt@qucam.com (SMT); prguez@ing.iac.es (PRG); vik.dhillon@sheffield.ac.uk (VSD)}, P. Rodr\'{i}guez-Gil$^{2,3}$  and V. S. Dhillon$^{1}$\\
$^{1}$Department of Physics and Astronomy, University of Sheffield, Sheffield, S3 7RH, UK\\
$^{2}$Isaac Newton Group of Telescopes, Apartado de Correos 321, Santa Cruz de La Palma, E-38700, Spain\\
$^{3}$Instituto de Astrof\'\i sica de Canarias, V\'\i a L\'actea, s/n, La Laguna, E-38205, Santa Cruz de Tenerife, Spain}
\begin{document}
\date{Accepted     . Received 2009}
\pagerange{\pageref{firstpage}--\pageref{lastpage}} \pubyear{2009}
\maketitle
\label{firstpage}
\begin{abstract}
We present high time-resolution spectroscopy of the eclipsing
cataclysmic variable SDSS\,J143317.78+101123.3 obtained with QUCAM2,
a high-speed/low-noise electron-multiplying CCD camera. \citet{Stu1433} measured the mass of the secondary star
in SDSS\,J143317.78+101123.3 using a light-curve fitting technique and obtained a value
of $M_2=0.060 \pm 0.003$ M$_\odot$, making it one of the three first bona-fide detections
of a brown-dwarf mass donor in a cataclysmic variable. In this paper
we present a dynamical measurement supporting this
important result. We measured the radial-velocity semi-amplitude of
the white dwarf from the motion of
the wings of the H$\alpha$ emission line and obtained a figure of $K_1=34 \pm 4$ km s$^{-1}$,
in excellent agreement with the value of
$K_1=35 \pm 2$ km s$^{-1}$ predicted by Littlefair et al.'s model.

\end{abstract}
\begin{keywords}
binaries: close -- binaries: eclipsing -- stars: dwarf novae -- novae, cataclysmic variables -- stars: individual : SDSS\,J143317.78+101123.3
\end{keywords}
\section{Introduction}

Cataclysmic variables (CVs) are close binary stars in which a
white-dwarf (WD) primary accretes material
from an accretion disc fed by a red-dwarf secondary (see \cite{HellierBook} and \cite{Warner} for reviews).
Theory predicts that the mass transfer from the secondary to the
primary is accompanied by a shortening of
the orbital period. This is believed to continue until the secondary
star drops below the hydrogen-burning limit
($\sim0.08 M_{\odot}$) and becomes a degenerate, brown-dwarf like
object, at which point the period begins
to increase. Models show that up to 70\% of all CVs in the Galaxy
should have brown-dwarf secondary stars
at the present time, yet not a single such object had been positively
identified until the discovery by \citet{Science1035} of a secondary star of mass $M_2=0.052\pm0.002 M_{\odot}$ in the CV SDSS\,J103533.03+055158.4.
Since then, two more brown-dwarf mass
donors have been discovered by \citet{Stu1433} SDSS\,J150137.22+550123.3 and SDSS\,J150722.30+523039.8 and it now
appears as if the missing population of
post-period minimum CVs has finally been identified.

The results of \citet{Stu1433} and \citet{Science1035}
rely on an eclipse light-curve fitting technique
applied to broad-band photometric data. The assumptions underlying
this technique appear to be robust
(see \cite{Stu1433} for a discussion), but the results are
of such significance that it is important
they are independently verified. In this paper we present
time-resolved spectroscopy of the CV SDSS\,J143317.78+101123.3 (hereafter SDSSJ1433),
obtained with the aim of measuring the radial-velocity semi-amplitude
of the white dwarf ($K_1$) and comparing
this direct, dynamical measurement of $K_1$ with that predicted by the
light-curve model of \citet{Stu1433}.
The observations were obtained with an electron-multiplying CCD (EMCCD), to the best of
our knowledge the first time that such a device has been used for astronomical spectroscopy.

\section{Observations and Reduction\label{sec-obs}}
SDSSJ1433 is a challenging target as it is faint ($g^{\prime}=18.5$) and has a short period ($P$=78.1 min). This means that spectra taken with a conventional CCD would be swamped by read noise and for this reason we chose to use an Electron Multiplying CCD \citep{Mackay}. These devices are able to amplify the signal to such an extent that it renders the readout noise negligible.
We observed SDSSJ1433 using QUCAM2\footnote{http://www.ing.iac.es/Engineering/detectors/qucam2.html}, an EMCCD-based camera mounted on the red arm of the ISIS spectrograph\footnote{http://www.ing.iac.es/Astronomy/instruments/isis/index.html} on the 4.2m William Herschel Telescope, La Palma. QUCAM2 employs a 1k $\times$ 1k  CCD201 detector manufactured by E2V which has a frame-transfer architecture, providing a dead time of only 12\,ms between each frame. With the low predicted $K_1$ it was decided to use the highest dispersion grating available, the R1200R. This provided a wavelength range of 6480--6700\AA{ }at a spectral resolution of 32 km s$^{-1}$ (0.7\AA) with a slit width of 1\arcsec.

On 2008 April 16, we obtained 628 spectra, each of 30s exposure time. The moon was full, the seeing was approximately 0.7--1\arcsec\ and the sky transparency was variable. A nearby comparison star was also placed on the slit to correct for slit losses and changes in transparency. Arc spectra were taken every 40 minutes for wavelength calibration. No flux standard was observed. This is justified because there was intermittent cloud, we only observed over a narrow wavelength range and we were only interested in extracting velocities from the data.

The spectra were extracted using a simple spatial bin across three windows within each frame covering the target, the reference star and the large area of sky in between.
The sky-subtracted spectra were cast into 40 phase bins using the ephemeris of \citet{Stu1433}, and then corrected for slit losses and transparency variations by dividing by the integrated flux in the comparison star spectra.
\section{Results \label{sec-results}}
\subsection{Averaged spectrum}
\begin{figure}
\begin{center}
\includegraphics[width=0.45\textwidth]{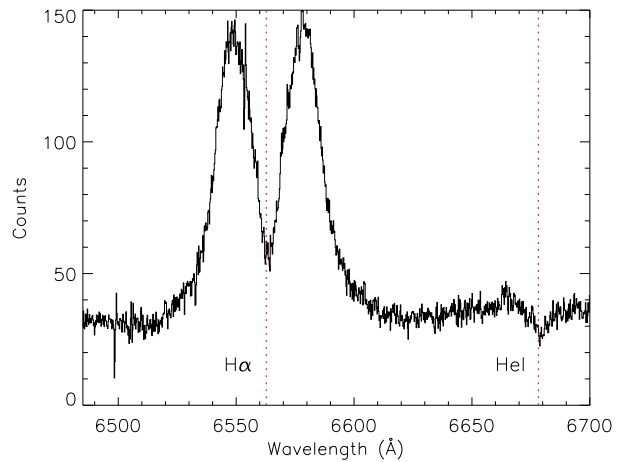}
\caption{The average spectrum of SDSSJ1433, obtained by combining all 628 spectra.}
\label{fig:averagespectrum}
\end{center}
\end{figure}

Figure~\ref{fig:averagespectrum} shows the average spectrum of SDSSJ1433. The most prominent feature is the broad, double-peaked emission line of H$\alpha$. The equivalent and velocity widths of the line are given in Table~\ref{results_table}. A weaker feature due to He\,{\sc i} $\lambda$6678 is also visible. The average spectrum is typical of other eclipsing dwarf novae below the period gap, e.g. WZ Sge \citep{Skidmore}.

\begin{table}
\begin{center}
\caption{Full-width at zero intensity, full-width at half maximum, peak-to-peak separation and equivalent width of the H$\alpha$ emission line shown in Figure~\ref{fig:averagespectrum}. The radial velocity semi-amplitude of the WD and the systemic velocity derived in Sections \ref{sec-diag} and \ref{light-centre} are also listed.}
\begin{tabular}{  l  l  }
    \hline
    FWZI & 5000 $\pm$ 500 km/s \\
    FWHM & 2200 $\pm$ 200 km/s \\
    Peak Separation & 1300 $\pm$ 200 km/s \\
    Equivalent Width & 147.2\ $\pm$ 0.5 \AA \\
    $K_1$ & 34 $\pm$ 4 km/s \\
    $\gamma$ & 75 $\pm$ 10 km/s \\
    \hline
\end{tabular}
\label{results_table}
\end{center}
\end{table}

\subsection{Continuum and H$\alpha$ light curves \label{sec-lc}}
\begin{figure}
\includegraphics[width=0.45\textwidth]{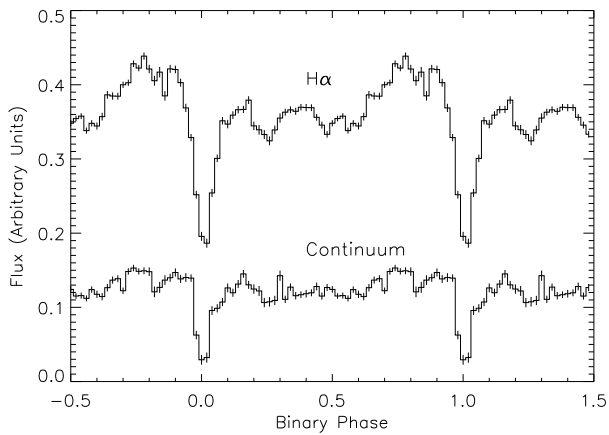}
\caption{Light curves of the continuum only and the H$\alpha$ emission line.}
\label{fig:lightcurves}
\end{figure}

We computed the light curve of the continuum by summing the flux in  line-free portions of each spectrum.  We then fitted and subtracted the continuum of each spectrum, and summed the residual flux in the H$\alpha$ emission line. The resulting light curves are shown in  Figure~\ref{fig:lightcurves}.

Figure~\ref{fig:lightcurves} shows that H$\alpha$ experiences broader eclipses than the continuum. The eclipse depth for H$\alpha$ is $\sim 50$ \%, while for the continuum it is $\sim 75$ \%. This is consistent with H$\alpha$ being emitted in the optically thin outer parts of the accretion disc and the hotter, optically thick, inner parts of the disc being responsible for the majority of the continuum emission. This picture is confirmed by analysing the light curves of the wings and core of the line separately, which show a deeper eclipse in the former than the latter. An orbital hump prior to eclipse is also apparent in the H$\alpha$ light curve of figure~\ref{fig:lightcurves}, caused by the changing aspect of emission from the bright spot.

\subsection{Trailed spectra and Doppler tomogram}
\begin{figure}
\begin{center}
\includegraphics[width=0.45\textwidth]{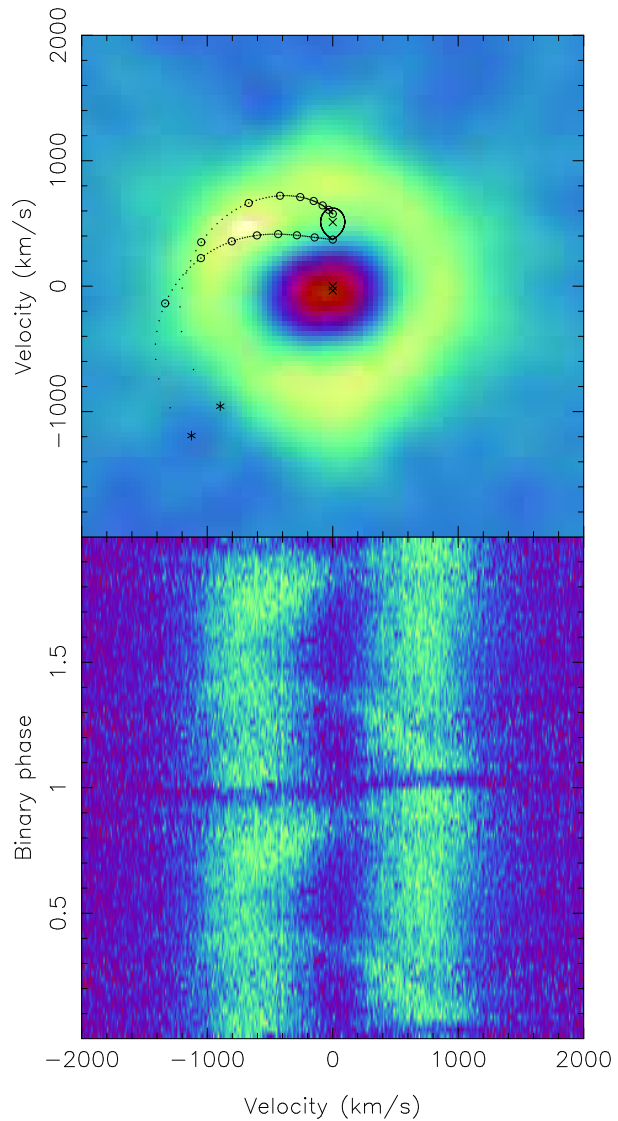}
\caption{\textit{Bottom:} the H$\alpha$ trailed spectrum, with two cycles plotted for clarity.
\textit{Top:} the Doppler map of H$\alpha$.  The three crosses  represent the centre of mass of the secondary star (upper cross), the system (middle cross) and the WD (lower cross). The Roche lobe of the secondary star, the predicted gas stream trajectory and the Keplerian velocity of the disc along the stream are also shown, calculated assuming the system parameters given in \citet{Stu1433}.
The circular tick marks represent steps of 0.1 L$_1$ (the inner Lagrangian distance) towards the WD.  The asterisks represent the points of closest approach to the WD.}
\label{fig:trail}
\end{center}
\end{figure}

The phase-binned, continuum-subtracted H$\alpha$ profile is shown as a trail in the lower panel of Figure~\ref{fig:trail}. The primary eclipse around phase 0 is clearly seen, as is the rotational disturbance where the blue peak of the emission line is eclipsed prior to the red peak. The bright spot is clearly visible as an S-wave moving between the two peaks. The orbital modulation of the wings of the emission line can arguably just be made out. There is also evidence for a \textquoteleft shadow\textquoteright\ on the blue edge of the S-wave around phase 0.25 which appears remarkably similar to the models of gas stream overflow computed by \citet{HellierGas}.

We computed the Doppler tomogram of the H$\alpha$ trail using Fourier-filtered back projection (see \cite{Marsh2001} for a review). As expected, the dominant feature is the ring of emission representing the accretion disc which appears to be centred on the expected position of the WD. The bright spot lies on this ring at a velocity intermediate between the free-fall velocity of the gas stream and the Keplerian disc velocity along its path, as observed in some other CVs, e.g. U Gem \citep{MarshUG} and WZ Sge \citep{WZSgt}. There is no evidence for emission from the secondary star, nor for asymmetries in the disc emission due to, for example, spiral shocks.

\subsection{The diagnostic diagram \label{sec-diag}}
The WD primary star orbits with velocity $K_1$ around the centre of mass of the system, resulting in a periodic Doppler shift of any light it emits.  If we imagine the simplified case of an ideal accretion disc centred on the WD with a perfectly symmetric brightness distribution, then the velocity shift of the emission lines generated in the disc will be subject to the same modulation as the WD. If we measure the line centroids and plot them as a function of orbital phase then we should see a sine wave corresponding to the orbital motion of the WD. More realistically, however, if we include the light emitted by the bright spot, we can see that calculated line centroids will, rather than precisely following the WD, be perturbed in the direction of the bright spot. The resultant radial velocity (RV) curve will not only have an excessive amplitude but will also be offset in phase with respect to the true motion of the WD. It is thus important to exclude the bright spot contribution by only examining the light emitted in the wings of the line profile. These emissions correspond to material orbiting within the disc at small radii from the primary star, where they are only minimally contaminated by the bright spot. If the RV curve that we obtain using this technique has the correct phase offset relative to the primary eclipse then we can be confident of an accurate semi-amplitude measurement.

\begin{figure}
\begin{center}
\includegraphics[width=0.45\textwidth]{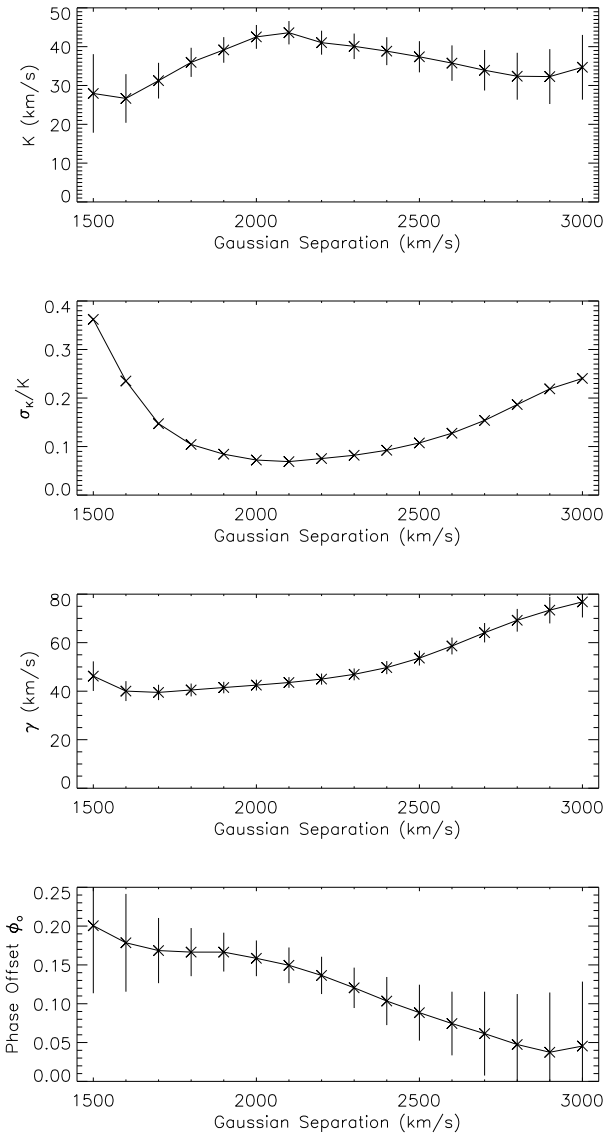}
\caption{The diagnostic diagram for H$\alpha$ -- see text for details.}
\label{fig:diagnostic}
\end{center}
\end{figure}

\begin{figure}
\includegraphics[width=0.450\textwidth]{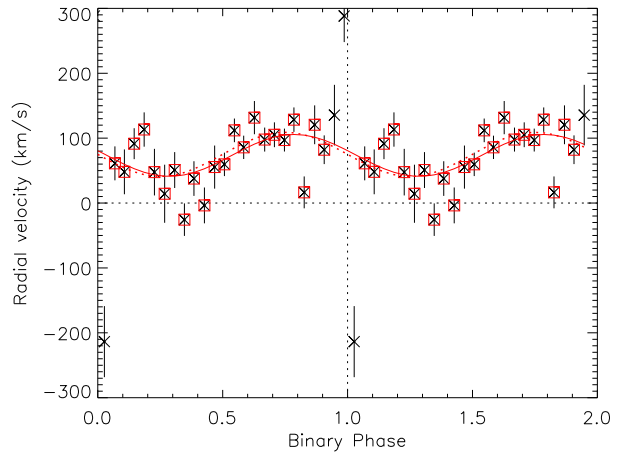}
\caption{The measured RV data for a Gaussian separation of 2900 km s$^{-1}$ and the resulting  sine curve fit (solid line). Only data points marked by a square were included in the fit, i.e. the rotational disturbance was not taken into account. The dashed line  shows the sine curve corresponding to the motion of the WD as derived from the light centre diagram presented in Figure~\ref{fig:lightcentre}.}
\label{fig:RVcurve}
\end{figure}

To measure the radial velocities of the wings of the H$\alpha$ emission line we used the double-Gaussian technique of \citet{schneider+young}. The best signal-to-noise ratio was obtained  using a Gaussian width of 350 km~s$^{-1}$ and the separation of the Gaussians was then varied between 1500 and 3000 km~s$^{-1}$ in steps of 100 km~s$^{-1}$ so as to explore annular regions of the accretion disc of decreasing radii. At each  Gaussian separation the resulting radial velocities were fitted with a sine function. The phase offset ($\phi_0$), semi-amplitude ($K$), systemic velocity ($\gamma$) and fractional error in the amplitude ($\sigma_K / K$) of the sine fits were then plotted against the Gaussian separation as a diagnostic diagram \citep{Shafter}, as shown in Figure~\ref{fig:diagnostic}. An example of the RV data and sine fit is shown in Figure~\ref{fig:RVcurve}, for a Gaussian separation of 2900 km s$^{-1}$.

The standard way of obtaining  $K_1$ from a diagnostic diagram is to use the value corresponding to where the $\sigma_K / K$ curve is at a minimum. If we do this we find a value of $K_1=44$km~s$^{-1}$. However, it can be seen that the corresponding value of $\phi_0$ is non-zero and declines towards zero at higher Gaussian separations. As pointed out by \citet{Marsh88}, this highlights the main drawback of diagnostic diagrams -- the derived $K_1$-value is noise dependent. The best solution is to use those data points corresponding to the highest Gaussian separations to extrapolate to a $K_1$-value whose corresponding $\phi_0=0$. This requires the construction of a \textquoteleft Light Centre\textquoteright\ diagram \citep{Marsh88}.

\subsection{The light centre method \label{light-centre}}
\begin{figure}
\begin{center}
\includegraphics[width=0.35\textwidth]{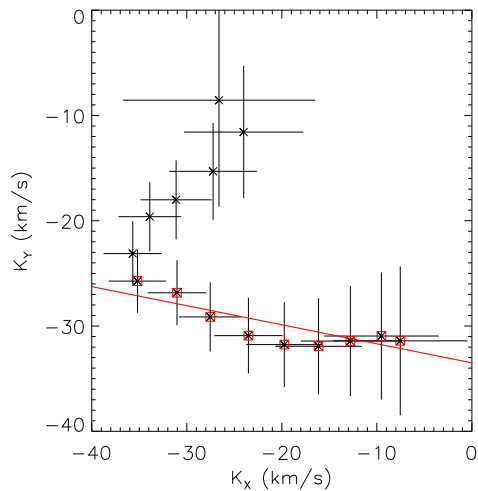}
\caption{The light centre diagram for H$\alpha$, where $-K \sin\phi_0$ is plotted on the abscissa and $-K \cos\phi_0$ on the ordinate. A linear fit to the points marked by squares is also shown, extrapolated to where it intercepts the $y$-axis. The smallest Gaussian separation corresponds to the top-most point and the largest to the right-most point. }
\label{fig:lightcentre}
\end{center}
\end{figure}

At no point on the diagnostic diagram does the sine fit have a zero phase offset relative to the WD, although at the extremes of the H$\alpha$ wings it gets close. The light centre method, described by \citet{Marsh88}, offers a way of extrapolating the data to estimate what the amplitude of the RV curve would be at $\phi_0 = 0$. Just as the Doppler map is a view of an emission line transformed into velocity space, so the light centre diagram, shown in Figure~\ref{fig:lightcentre}, is a velocity space projection of the RV data plotted in the diagnostic diagram.

The radial velocity of the WD can be determined from a light centre diagram by extrapolating a linear fit to the RV data points and reading off the $y$-axis intercept. To avoid contamination from the bright spot we only included the nine largest Gaussian separations in the linear fit. The resulting intercept is $K_1 = 34 \pm 4$ km s$^{-1}$, the error quoted corresponding to a combination of statistical error and systematic error due to the exact choice of which points to exclude from the linear fit. The RV curve corresponding to this value of $K_1$ and $\phi_0 = 0$ is shown as the dashed curve in  Figure~\ref{fig:RVcurve}.

We are confident that our measurement of $K_1$ accurately reflects the motion of the WD in
SDSSJ1433 for the following reasons. First, the bottom panel of figure~\ref{fig:diagnostic}
clearly shows a trend to zero phase offset at large Gaussian separations, as one would expect
if one is measuring the real radial velocity of the WD in combination with a steadily
decreasing contribution from the bright spot. In such instances, the light centre technique
has been proven to give reliable results (e.g. IP Peg, \citealt{Marsh88}; U Sco, \citealt{Thoroughgood01}).
Second, the top panel of figure~\ref{fig:diagnostic} shows that the radial velocity semi-amplitude
is remarkably independent of the Gaussian separation, indicating that our systematic errors are
indeed small.

\section{Conclusions \label{sec-conclusion}}
We have measured the radial velocity amplitude, $K_1 = 34 \pm 4$~km s$^{-1}$, of the WD in the short-period CV SDSSJ1433 by analysing the RV motion of the extreme wings of the H$\alpha$ emission line. In order to account for the velocity contamination from the conspicuous emission of the bright spot, the light centre technique was used. The measured value of $K_1$ is consistent with the predicted value of $K_1=35 \pm 2$ km s$^{-1}$  from the light-curve fitting technique of \citet{Stu1433}. Our result  therefore supports the validity and accuracy of the purely photometric technique of measuring masses and argues in favour of the presence of a brown-dwarf donor in SDSSJ1433, and by implication in SDSSJ1501 and SDSSJ1507 also.
The results presented in this paper also demonstrate that the combination of a large-aperture telescope, an intermediate-resolution spectrograph and  an EMCCD is uniquely capable of tackling this type of photon-starved observation.

\section*{Acknowledgments}
We thank Tom Marsh for the use of his \texttt{MOLLY} and \texttt{TRAILER} programs and Stuart Littlefair for useful discussions.
The observations were made with the William Herschel
Telescope operated on the island of La Palma by the
Isaac Newton Group in the Spanish Observatorio del Roque
de los Muchachos of the Instituto de Astrof\'\i sica de Canarias.

\appendix
\bsp
\label{lastpage}
\end{document}